\documentstyle[aps,twocolumn,prl,epsfig]{revtex}

\begin{document}

\title{
Quantum Monte Carlo study of a nonmagnetic impurity in the
two-dimensional Hubbard model
}

\author{ N.\ Bulut }

\address{
Department of Physics, Ko\c{c} University, Sariyer, 34450 Istanbul, Turkey \\
nbulut@ku.edu.tr
}

%\date{\today}
%\draft

\twocolumn[\hsize\textwidth\columnwidth\hsize\csname@twocolumnfalse\endcsname
\maketitle

\begin{abstract}
In order to investigate the effects of nonmagnetic impurities in
strongly correlated systems, Quantum Monte Carlo (QMC) simulations
have been carried out for the doped two-dimensional Hubbard model
with one nonmagnetic impurity. Using a bare impurity potential
which is onsite and attractive, magnetic and single-particle
properties have been calculated. The QMC results show that giant
oscillations develop in the Knight shift response around the
impurity site due to the short-range antiferromagnetic
correlations. These results are useful for interpreting the NMR
data on Li and Zn substituted layered cuprates.

\pacs{PACS Numbers:}

%Keywords:

\end{abstract}
]

\newpage

\section{Introduction}
The substitution of impurities into strongly-correlated systems is
a useful probe of the intrinsic electronic correlations. For
instance, in the layered cuprates it has been found that the
substitution of nonmagnetic impurities in place of Cu in the
CuO$_2$ planes strongly influences the superconducting \cite{Xiao}
and the magnetic correlations
\cite{Mahajan94,Bobroff,Mendels,Mahajan2000,MacFarlane,Julien}.
These experiments have provided valuable insight into the
interplay of the magnetic, superconducting and the density
correlations in these materials. In particular, the NMR
experiments find that when a uniform magnetic field is applied,
the electronic spins in the CuO$_2$ planes in
YBa$_2$Cu$_3$O$_{6+x}$ polarize to form an oscillatory pattern
which decays away from the impurities
\cite{Mahajan94,Bobroff,Mendels,Mahajan2000,MacFarlane,Julien}.
These are observed as giant oscillations in the Knight shift
response of the system. Furthermore, the Knight shifts for nuclei
close to the impurity have a perfect Curie-Weiss temperature
dependence. The origin of these behaviors is an important
question.

Theoretically, the problem of a nonmagnetic impurity in strongly
correlated systems such as the two-dimensional (2D) $t$-$J$ and
Hubbard models have been studied using various many-body
techniques. The single-particle properties of an impurity in 2D
$t$-$J$ clusters were studied with the exact diagonalization
technique \cite{Poilblanc,Ziegler96}. The magnetic and
single-particle correlations around a nonmagnetic impurity in the
doped 2D Hubbard model were studied with the the slave boson and
the Quantum Monte Carlo (QMC) techniques \cite{Ziegler98}. These
calculations have found Friedel oscillations in the electron
density. There are also RPA type of calculations of the Knight
shifts for a nonmagnetic impurity in the 2D Hubbad model
\cite{Bulut2000,Bulut2001,Ohashi2002}, which suggest that the
locally enhanced antiferromagnetic (AF) correlations around the
impurity are responsible for the giant oscillations and the
Curie-Weiss behavior observed in the Knight shift measurements.

In this paper, the magnetic and single-particle correlations
around one nonmagnetic impurity in the Hubbard lattice will be
studied with the QMC technique. For this purpose, the
determinantal QMC algorithm introduced by Blankenbecler, Scalapino
and Sugar \cite{BSS} and described by Ref. \cite{White89} will be
used. The emphasis in this paper will be on the Knight shift
response of the system in the metallic state near half-filling.
The QMC data show that as short-range AF fluctuations grow, giant
oscillations develop in the Knight shift response. Here, the
real-space structure of these oscillations as well as the pattern
in the electron density will be discussed. In addition, the
effective impurity potential $V_{eff}$ will be extracted from the
QMC data on the single-particle Green's function. It will be seen
that $V_{eff}$ is weakly attractive at the sites neighboring the
impurity. Because of the ``fermion sign problem", the QMC
calculations cannot be performed at sufficiently low temperatures
where direct comparisons with the NMR data can be made. Instead,
the QMC data will be compared with the RPA-type of calculations
used for fitting the Knight shift data \cite{Bulut2001}. It will
be seen that the real-space patterns of the Knight shift, the
electron density and the effective impurity interaction obtained
with the QMC technique are in good agreement with the results
found in the RPA analysis of the Knight shift data. These
emphasize the role of the AF fluctuations in producing the unusual
Knight shift data on Li and Zn substituted YBa$_2$Cu$_3$O$_{6+x}$.
An important feature of these calculations is that here the system
is always in the paramagnetic state, and no AF droplets have
formed around the impurity.

In the following, first the model used for a nonmagnetic impurity
in the Hubbard model will be discussed. Then, the numerical data
on the Knight shift and the local electron density will be
presented. Next, these results will be compared with the previous
RPA analysis of the Knight shift measurements. Finally, the QMC
data on the effective impurity potential will be presented.

\section{Model}
The 2D Hubbard model is defined by
\begin{equation}
\label{Hubbard} H=-t\sum_{\langle i,j\rangle ,\sigma}
(c^{\dagger}_{i\sigma}c_{j\sigma}
+c^{\dagger}_{j\sigma}c_{i\sigma}) + U \sum_i n_{i\uparrow}
n_{i\downarrow} \nonumber -\mu \sum_{i} n_i
\end{equation}
where $t$ is the hopping matrix element, $U$ is the onsite Coulomb
repulsion and $\mu$ is the chemical potential. The electron
creation (annihilation) operator at site $i$ with spin $\sigma$ is
represented by $c_{i\sigma}^{\dagger}$ ($c_{i\sigma}$),
$n_{i\sigma}=c_{i\sigma}^{\dagger}c_{i\sigma}$ is the electron
occupation number with spin $\sigma$ and $n_i=n_{i\uparrow} +
n_{i\downarrow}$. In the following, numerical data will be
presented for $\langle n\rangle =0.875$, and $U=4t$ and $8t$.

A nonmagnetic impurity located at the origin will be modelled by
the onsite one-electron potential
\begin{equation}
\label{Vimp}
V_{imp} = V_0 n_0.
\end{equation}
The value of $V_0$ will be taken to be $-20t$, so that it is
greater than both $U$ and the bandwidth. For this value of $V_0$,
the electron occupation number of the impurity site is nearly two.
In addition, the local magnetic moment defined by
\begin{equation}
m({\bf r})=\sqrt{\langle (m^z({\bf r}))^2 \rangle},
\end{equation}
where $m^z({\bf r}_i)=n_{i\uparrow}-n_{i\downarrow}$, is close to
zero at the impurity site ${\bf r}=(0,0)$. So, for this value of
$V_0$, the impurity site is nearly doubly-occupied and
magnetically inert. Hence, one would expect that this model
captures some of the physical properties of nonmagnetic impurities
in strongly correlated systems, even though this form of the bare
impurity potential is a very simple one.

The Knight shift of the various nuclear sites is determined by the
magnetic susceptibility $\chi$ through the relation
\begin{equation}
\label{kr} k({\bf r}_i) = \sum_j \chi( {\bf r}_i, {\bf r}_j , i
\omega_m =0 ),
\end{equation}
where $j$ sums over the whole lattice \cite{Bulut2001}. For
instance, if the hyperfine coupling for the nuclear spin ${\bf
I}_i$ at site ${\bf r}_i$ is $A{\bf I}_i \cdot {\bf S}_i$, then
the corresponding Knight shift is given by $(\gamma_e/2\gamma_n) A
k({\bf r}_i)$, where $\gamma_e$ and $\gamma_n$ are the electronic
and the nuclear gyromagnetic ratios. Here, the staggered magnetic
susceptibility $\chi$ is defined by
\begin{equation}
\chi({\bf r}_i, {\bf r}_j , i\omega_m ) = \int_0^{\beta} d\tau \,
e^{i\omega_m \tau} \, \langle m_i^-(\tau) m_j^+(0) \rangle
\end{equation}
where $\omega_m=2m\pi T$ is the Bose Matsubara frequency, $m_i^+=
c^{\dagger}_{i\uparrow}c_{i\downarrow}$, $m_i^-=
c^{\dagger}_{i\downarrow}c_{i\uparrow}$ and $m_i^-(\tau) =
e^{H\tau} m_i^- e^{-H\tau}$. The expectation value $\langle
...\rangle$ is evaluated with respect to the hamiltonian of the
impure system, which is given by $H+V_{imp}$. Note that for the
translationally invariant pure system, $k({\bf r})$ reduces to the
uniform magnetic susceptibility $\chi_{pure}({\bf q}\rightarrow
0,i\omega_m=0)$. In addition to these, results on the local
magnetic moment $m({\bf r})$ will be shown. Note that the
site-dependent electron density $\langle n({\bf r})\rangle$ and
$m({\bf r})$ were previously calculated \cite{Ziegler98}. Here,
these quantities are reproduced for comparisons with $k({\bf r})$
and the effective impurity potential.

The transferred hyperfine coupling between the nuclear spin of a
$^7$Li impurity, $^7{\bf I}$, and the electronic spins in the
CuO$_2$ planes can be described by
\begin{equation}
C ^7{\bf I} \cdot \sum_{i=1}^4 {\bf S}_i,
\end{equation}
where $C$ is the hyperfine coupling constant and $i$ sums over the
four nearest-neighbor Cu sites of the $^7$Li impurity. In this
case, the Knight shift for $^7$Li is
\begin{equation}
^7K = {1 \over 2} \big( {\gamma_e \over ^7\gamma_n} \big) C 4
k({\bf r}=(1,0)),
\end{equation}
so that the temperature dependence of $^7K$ is determined by
$k({\bf r})$ evaluated at the nearest-neighbor site of the
impurity, ${\bf r}=(1,0)$. The measurements of the $^7$Li Knight
shift and of the $^{89}$Y Knight shift at the first and the second
nearest-neighbor Y sites of the impurity determine $k({\bf r})$ in
the vicinity of the impurity \cite{Mahajan94,Bobroff,Mahajan2000}.
The analysis of the $^7$Li and $^{89}$Y Knight shift data on Li
and Zn substituted YBa$_2$Cu$_3$O$_{6+x}$ concluded that $k({\bf
r})$ develops giant oscillations as $T$ decreases
\cite{Mahajan2000,Bulut2001}. The $T$-dependent line broadening of
the $^{63}$Cu NMR spectra in Zn substituted YBa$_2$Cu$_3$O$_{6.7}$
has been also attributed to the development of a staggered
polarization of the electronic spins, when a uniform magnetic
field is applied \cite{Julien}. The QMC data presented here show
that it is the AF correlations which are responsible for the
pattern in $k({\bf r})$.

In addition to the magnetic properties, the single-particle
properties around the impurity will be studied. The
single-particle Green's function is defined by
\begin{equation}
G({\bf r}_i,{\bf r}_j, i\omega_n ) = - \int_0^{\beta} d\tau \,
e^{i\omega_n\tau} \langle c_{i\sigma} (\tau) c_{j\sigma}^{\dagger}
(0) \rangle ,
\end{equation}
where $\omega_n = (2n+1)\pi T$ is a Fermi Matsubara frequency.
Here, first the site-dependent electron density $\langle n({\bf
r}) \rangle$ will be discussed. In order to understand the ${\bf
r}$ dependence of $\langle n({\bf r})\rangle$, the effective
impurity potential will be extracted from the QMC data on $G$ of
the impure system and $G_U$ of the pure Hubbard system. The
Dyson's equation relating $G$ and $G_U$ is
\begin{eqnarray}
G({\bf r}&&,{\bf r'}, i\omega_n) = G_U({\bf r},{\bf r'},
i\omega_n) \\
&&+ \sum_{{\bf r_1},{\bf r_2}} G_U({\bf r},{\bf r_1}, i\omega_n)
T({\bf r_1},{\bf r_2},i\omega_n) G({\bf r_2},{\bf r'}, i\omega_n)
\nonumber
\end{eqnarray}
where $T({\bf r},{\bf r'},i\omega_n)$ is the effective impurity
scattering matrix. Because of the Coulomb correlations, $T({\bf
r},{\bf r'},i\omega_n)$ becomes extended in real space
\cite{Ziegler96}. Here, the diagonal component of $T$ is defined
as the effective impurity potential,
\begin{equation}
V_{eff}({\bf r},i\omega_n) = T({\bf r},{\bf r},i\omega_n).
\end{equation}
In Section III.C, the spatial structure of $T({\bf r},{\bf
r'},i\omega_n)$ will be discussed and results on $V_{eff}({\bf
r},i\pi T)$ will be shown.

\section{Numerical Data}
\subsection{Knight shift}

The following data were obtained for electron filling $\langle
n\rangle = 0.875$ and an $8\times 8$ lattice with periodic
boundary conditions. First, results will be shown for the $U=4t$
case at temperatures between $0.25t$ and $1.0t$. In Fig.~1(a),
$k({\bf r})$ is plotted as a function of $r =|{\bf r}|$ in units
of the lattice spacing $a$. The impurity is located at the origin
at ${\bf r}=(0,0)$. For instance, in this notation, $r=1$
corresponds to $(\pm 1,0)$ and $(0,\pm 1)$ sites whereas
$r=\sqrt{2}$ corresponds to $(\pm 1,\pm 1)$ and $(\pm 1,\mp 1)$.
In addition, here $k({\bf r})$ is shown in units of $t^{-1}$ and
$t$ is taken to be unity.

In Fig.~1(a), it is seen that $k({\bf r})$ vanishes at the
impurity site. This is because the strong impurity potential
$V_0=-20t$ overcomes the electron-electron repulsion $U$ and binds
an up-spin and a down-spin electron at ${\bf r}=(0,0)$, so that
the impurity site becomes magnetically inert. In the vicinity of
the impurity, $k({\bf r})$ exhibits oscillations, which grow as
the temperature decreases. The horizontal long-dashed line in
Fig.~1(a) represents the value of $k({\bf r})$ for the pure system
at $T=0.25t$. Hence, away from the impurity, $k({\bf r})$ goes to
its value for the pure system, which is the uniform magnetic
susceptibility. Since these calculations are carried out at high
temperatures, the uniform susceptibility does not yet exhibit the
temperature-independent Pauli behavior. These data on $k({\bf r})$
show that the nonmagnetic impurity strongly influences the
zero-frequency magnetic correlations around it.

In Figure 1(b), results on $m({\bf r})$ are plotted in the same
way as in Fig. 1(a). Here, it is seen that, between $T=0.25t$ and
$1.0t$, $m({\bf r})$ is very weakly dependent on the temperature.
At ${\bf r}=(0,0)$, $m({\bf r})$ has a small value. Comparing with
Fig.~1(a), it is observed that the oscillations in $k({\bf r})$
are stronger than those in $m({\bf r})$. Note that both in $k({\bf
r})$ and $m({\bf r})$, the maximum occurs at ${\bf r}=(1,0)$. The
NMR experiments also find that $k({\bf r})$ has its maximum value
at ${\bf r}=(1,0)$.

In order to understand the origin of the oscillatory structure in
$k({\bf r})$, it is useful to consider the Fourier transform of
$\chi({\bf r},{\bf r'},i\omega_m)$ defined by
\begin{equation}
\chi({\bf q},{\bf q'},i\omega_m) = \sum_{{\bf r},{\bf r'}} \,
e^{-i({\bf q}\cdot {\bf r} - {\bf q'}\cdot{\bf r'})} \chi({\bf
r},{\bf r'},i\omega_m).
\end{equation}
This quantity is related to the Fourier transform $k({\bf q}) =
\sum_{\bf r} \, e^{-i{\bf q}\cdot {\bf r}} k({\bf r})$ through
\begin{equation}
k({\bf q}) = \chi({\bf q}, {\bf q'}=0, i\omega_m =0).
\end{equation}
In Fig. 2(a), $-k({\bf q})$ is plotted at different temperatures
as a function of ${\bf q}$ along various cuts in the Brillouin
zone. For comparison, in Fig.~2(b) the diagonal susceptibility
\begin{equation}
\chi({\bf q}) = \chi({\bf q},{\bf q},i\omega_m=0)
\end{equation}
versus ${\bf q}$ is plotted. In these figures, both $k({\bf q})$
and $\chi({\bf q})$ are plotted in units of $t^{-1}$. Here, it is
seen that $-k({\bf q})$ develops a sharp peak at ${\bf
q}=(\pi,\pi)$, which closely follows the development of the peak
in $\chi({\bf q})$ and, hence, of the AF correlations in the
system. This relation between $-k({\bf q})$ and the AF
correlations becomes more clear if the following RPA expression
for $\chi({\bf q},{\bf q'},i\omega_m)$ is considered
\cite{Bulut2000,Bulut2001},
\begin{eqnarray}
\chi({\bf q},{\bf q'},i\omega_m)&& = \chi_0({\bf q},{\bf
q'},i\omega_m)\\ &&+ \overline{U} \sum_{\bf q"} \chi({\bf q}, {\bf
q"}, i\omega_m) \chi_0({\bf q"},{\bf q'} , i\omega_m), \nonumber
\end{eqnarray}
where $\overline{U}$ is the effective irreducible vertex in the
particle-hole channel and $\chi_0({\bf q"},{\bf q'};i\omega_m)$ is
the susceptibility of the impure $U=0$ system. For ${\bf q'}=0$
and $i\omega_m=0$, one obtains
\begin{equation}
k({\bf q}) = k_0({\bf q}) + \overline{U} \sum_{\bf q"} \,
\chi({\bf q},{\bf q"} ,0) k_0({\bf q"}),
\end{equation}
where $k_0({\bf q})$ is for the impure $U=0$ system. This
expression shows that the peak in $k({\bf q})$ at ${\bf
q}=(\pi,\pi)$ is coupled to the AF fluctuations. The QMC data seen
in Figures 2(a) and (b) show that this is true for the real system
also and that it is the AF correlations which cause the
development of the giant oscillations in $k({\bf r})$.

At this point, it is necessary to discuss the physical meaning of
the ${\bf q}$ dependence of $k({\bf q})$. The QMC data and the
Knight shift experiments show that $k({\bf q})$ peaks at ${\bf
q}\sim (\pi,\pi)$. Since $k({\bf q})$ is equivalent to the
off-diagonal susceptibility $\chi({\bf q},{\bf
q'}=0,i\omega_m=0)$, this result means that the scattering of the
AF spin fluctuations with $\sim (\pi,\pi)$ momentum transfer is
one of the important effects of the nonmagnetic impurities. This
then gives information about the response of the system in the
magnetic channel to a perturbation in the density channel.

It is useful to compare the structure in $k(\bf r)$ with the
site-dependent electron density $\langle n({\bf r}) \rangle$.
Figure 3 shows $\langle n({\bf r})\rangle$ versus $r$ for the same
parameters as in Figures 1 and 2. At the impurity site ${\bf
r}=(0,0)$, the electron density is nearly equal to two and it is
not included in this graph. There are oscillations in $\langle
n({\bf r})\rangle$ which decay as one moves away from the
impurity. For sites close to the impurity and especially for ${\bf
r}=(1,0)$ and $(1,1)$, $\langle n({\bf r})\rangle$ increases as
$T$ decreases.

Figures 4(a) through (d) show QMC data on $k({\bf r})$, $-k({\bf
q})$, $\chi({\bf q})$ and $\langle n({\bf r})\rangle$ for the
stronger coupling case of $U=8t$. These data are plotted in the
same way as for $U=4t$ above. The structures in $k({\bf r})$,
$k({\bf q})$, $\chi({\bf q})$ are similar to what have been seen
for $U=4t$, only the amplitudes differ. When $U$ increases to
$8t$, the impurity site remains doubly occupied, however the
structure in $\langle n({\bf r})\rangle$ changes. Compared to the
$U=4t$ case, here $\langle n({\bf r}=(1,0))\rangle$ has a bigger
enhancement over the background. As it will be seen in Section
III.C, this has to do with the fact that $V_{eff}({\bf r},i\pi T)$
at ${\bf r}=(1,0)$ is more attractive for $U=8t$.

These calculations for $U=4t$ and $8t$ were repeated for an
electron filling of 0.94. In this case, the system has
longer-range AF fluctuations. Consequently, $k({\bf r})$ is more
enhanced and has stronger oscillations. However, the structures in
$k({\bf r})$, $k({\bf q})$, $\chi({\bf q})$ and $\langle n({\bf
r})\rangle$ are similar to what have been shown for the $\langle
n\rangle =0.875$ case. Because of space limitations these results
will not be presented here.

\subsection{Comparison with the RPA analysis of the Knight shift
data}

The QMC data on the Knight shifts are at high temperatures to be
directly compared with the experimental data. However, the
real-space structure of $k({\bf r})$ and $\langle n({\bf
r})\rangle $ calculated with QMC are in good agreement with the
results of the RPA analysis of the experimental data
\cite{Bulut2001}. In the RPA approach, an effective impurity
potential which is extended in real space and independent of
frequency,
\begin{equation}
V_{eff} = V_0 n_0 + V_1 \sum_{i=1}^4 n_i,
\label{Veff}
\end{equation}
was used to model the nonmagnetic impurity. The onsite component
$V_0$ was set to a large attractive value, and the extended
component $V_1$, which is acting at the four sites neighboring the
impurity, was used as a fitting parameter. Both $V_0$ and $V_1$
were taken to be real valued. Then, the magnetic susceptibility
and $k({\bf r})$ were calculated by using an RPA-like approach for
treating the Coulomb correlations. The magnetic susceptibility
$\chi_0$ of the impure $U=0$ system entering the RPA expression
was calculated by including the self-energy and the vertex
corrections due to the impurity scattering. It was found that for
an effective quasiparticle bandwidth of 1eV, the $^7$Li Knight
shift data on the optimally doped YBa$_2$Cu$_3$O$_{6+x}$ can be
fitted over a wide temperature range by using $V_1=-0.15t$.

In this approach, the magnitudes of the quantities such as $k({\bf
r})$ or the value of $V_1$ used for fitting the data depends on
the effective bandwidth, and it is difficult to know what the
exact value of the effective bandwidth should be in such a model.
In addition, there are uncertainties in the values of the
hyperfine couplings. For these reasons, what should really be
compared with is the pattern of the oscillations in $k({\bf r})$
and $\langle n({\bf r})\rangle$, and not so much the magnitudes.

In Fig. 5(a), $k({\bf r})$ obtained by fitting the $^7$Li Knight
shift, $^7K$, in optimally doped YBa$_2$Cu$_3$O$_{6+x}$ with
$^7$Li impurities is shown at 100K and 400K. In this calculation,
the $^7$Li hyperfine coupling was taken to be $1.8\times
10^{-20}$erg, corresponding to 0.85kOe/$\mu_B$. It is useful to
compare $k({\bf r})$ seen in Fig. 5(a) with the QMC data presented
in Figures 1(a) and 4(a). Here, it is seen that as $T$ decreases,
the QMC data develop real-space structure which is similar to the
results of the RPA analysis.

It is difficult to compare the experimental data with the QMC
calculations directly, but it would still be useful to discuss the
temperature range where the QMC calculations have been performed.
The single-band parameters which are considered to be appropriate
for the cuprates are $t\sim 0.45$eV and $U$ of order $12t$
\cite{Hybertson}. It is difficult to reach low temperatures with
large $U/t$. Hence, here calculations have been carried out for
$U=8t$ and $4t$. For $U=8t$, QMC data have been shown down to
$T=0.33t$. For this value of $U$, the magnetic exchange $J\sim
4t^2/U$ is $0.5t$, and, hence, $T=0.33t$ corresponds to two-thirds
of $J$, which is too high to make comparisons with the
experiments. For the intermediate coupling $U=4t$ case, the lowest
$T$ where QMC data are shown is $0.25t$. In this case, the
magnetic exchange $J\sim t$, and $T=0.25t$ corresponds to $J/4$.
In fact, the QMC data on $k({\bf r})$ for $U=4t$ and $T=0.25t$,
which are shown by the filled circles in Fig. 1(a), compare well
with the RPA fits to the experimental data at 400K, shown by the
empty circles in Fig. 5(a). This is encouraging, however, the
expression $J\sim 4t^2/U$ for the magnetic exchange is valid in
the strong coupling limit. So, it is difficult to compare the QMC
data directly with the experiments. On the other hand, studying
how $k({\bf r})$ evolves with $U/t$ and $T/t$ shows what to expect
for the Knight shifts at low temperatures in the strong coupling
limit.

Within the same RPA framework and using an effective bandwidth of
1eV, the Knight shift data on $^7$Li and $^{89}$Y were also fitted
for the underdoped YBa$_2$Cu$_3$O$_{6+x}$ with nonmagnetic
impurities. In this case, it was found that, as $T$ decreases from
400K to 100K, $k({\bf r}=(1,0))$ increases from $0.8 t^{-1}$ to
$2.1 t^{-1}$, whereas $k({\bf r}=(1,1))$ goes from $0.04 t^{-1}$
to $-0.8 t^{-1}$. This clearly demonstrates the severe effect of
the nonmagnetic impurity on the zero-frequency magnetic
correlations in the underdoped cuprates. At the temperatures where
they are performed, the QMC calculations also find that the
oscillations in $k({\bf r})$ are filling dependent. For example,
for $U=4t$ and $T=0.25t$, $k({\bf r}=(1,0))$ increases by about
10\%, when $\langle n\rangle$ varies from 0.875 to 0.94. On the
other hand, if $\langle n\rangle$ goes from 0.875 to 0.80, then
$k({\bf r}=(1,0))$ decreases by about 30\%. In this paper, QMC
results are shown for $\langle n\rangle =0.875$, because, for this
value of the filling and at the temperatures where these
calculations are carried out, the system has short-range AF
correlations which are not weak, and their effects are visible.

Next, comparisons are made for the local electron density. Fig.
5(b) shows $\langle n({\bf r})\rangle$ versus ${\bf r}$ obtained
from the same RPA analysis at 100K for different values of $V_1$.
The oscillations seen in this figure are stronger than in the QMC
data, which were obtained at higher temperatures. Here, it is seen
that as $V_1$ becomes more attractive, the electron density at
${\bf r}=(1,0)$ increases. Comparison with the QMC data in Figures
3 and 4(d) implies that the effective impurity potential at ${\bf
r}=(1,0)$ is more attractive for $U=8t$ than for $4t$. In the next
subsection, it will be seen that this is indeed the case.

\subsection{Effective impurity interaction}
In order to understand the spatial structure of $\langle n({\bf
r})\rangle$, it is necessary to discuss the effective scattering
matrix $T({\bf r},{\bf r'},i\omega_n)$, which can be obtained from
\begin{equation}
{\bf T}(i\omega_n) = {\bf G}_U^{-1}(i\omega_n) - {\bf
G}^{-1}(i\omega_n).
\end{equation}
Here, ${\bf T}(i\omega_n)$ is a matrix of which $(i,j)$'th element
is $T({\bf r}_i,{\bf r}_j,i\omega_n)$, ${\bf T}^{-1}$ is the
inverse of ${\bf T}$, and similarly for ${\bf G}$ and ${\bf G}_U$.
In the following, QMC results will be shown for the lowest
Matsubara frequency $\omega_n=\pi T$. For ${\bf r}$ and ${\bf r'}$
more than one or two lattice spacings away from the impurity,
$T({\bf r},{\bf r'},i\pi T)$ is small in the parameter regime the
QMC calculations were performed. The diagonal terms of the
scattering matrix $T$ are represented by $V_{eff}({\bf
r},i\omega_n)$. At the impurity site ${\bf r}=(0,0)$,
$V_{eff}({\bf r},i\pi T)$ is reduced from its bare value of $-20t$
due to the Coulomb correlations, and it also has an imaginary
part. For $U=4t$ and $T=0.25t$, $V_{eff}({\bf r},i\pi T)$ at ${\bf
r}=(0,0)$ is $-18t+i\,0.6t$, while for $U=8t$ and $T=0.5t$, it is
$-15t+i\,2t$. Away from the impurity site, $V_{eff}({\bf r},i\pi
T)$ is largest at $(\pm 1,0)$ and $(0,\pm 1)$. In addition, the
effective scattering matrix has large matrix elements between the
impurity site and the sites neighboring it. Hence, the presence of
the impurity renormalizes the bare hopping matrix elements between
the impurity and its neighbors. The off-diagonal terms such as
$T({\bf r}=(1,0),{\bf r'}=(0,1),i\pi T)$ are small compared to
$V_{eff}({\bf r}=(1,0),i\pi T)$.

Fig. 6(a) shows the real-space structure of the real and the
imaginary parts of $V_{eff}({\bf r},i\pi T)$ for $U=4t$ and
$T=0.25t$, and Fig. 6(b) shows the temperature dependence of
$V_{eff}({\bf r}=(1,0),i\pi T)$. Here, it is seen that $V_{eff}$
at ${\bf r}=(1,0)$ is weakly attractive. The results for $U=8t$
are shown in Figures 7(a) and (b). In this case, $V_{eff}({\bf
r}=(1,0),i\pi T)$ has a bigger value. At $T=0.5t$, its real and
imaginary parts are of order $-t/4$. This is why $\langle n({\bf
r}=(1,0)\rangle$ for $U=8t$ is more enhanced over the background
than for $4t$.

In previous analysis of the NMR data \cite{Bulut2001}, an RPA type
of approximation was used for treating the Coulomb correlations
and the effective impurity potential was taken to be extended in
space with the form given by Eq.~(\ref{Veff}), but it was assumed
that $V_{eff}$ is real-valued and independent of frequency. Within
this model, the fitting of the Knight shift data required that
$V_{eff}$ is weakly attractive at the sites neighboring the
impurity. The QMC data presented in Figures 6 and 7 show that the
real part of $V_{eff}({\bf r},i\pi T)$ at ${\bf r}=(1,0)$ is
attractive, which is in agreement with the RPA fitting of the
data. The QMC calculations also find that $T({\bf r},{\bf
r'},i\omega_n)$ depends on frequency, and, in addition, it has an
imaginary component. Since ${\rm Im}\,T({\bf r},{\bf
r'},i\omega_n)$ is odd in $\omega_n$, ${\rm Im}\,T({\bf r},{\bf
r'},i\pi T)$ should vanish as the temperature goes to zero. In
order to investigate the effects of $V_{eff}$ having an imaginary
component, the RPA calculations were repeated using
$V_1(i\omega_n) = -0.15t\,(1+i\,{\rm sign}(\omega_n))$ instead of
$V_1=-0.15t$ in Eq.~(\ref{Veff}), and keeping all other parameters
the same in the calculations. It was found that the imaginary
component of $V_1$ affects the magnitudes of $k({\bf r})$ and
$\langle n({\bf r})\rangle$, but the patterns in these quantities
remain similar. These comparisons with the QMC data on $T({\bf
r},{\bf r'},i\omega_n)$ support the form of $V_{eff}$,
Eq.~(\ref{Veff}), used in the RPA analysis of the Knight shift
experiments.

\section{Conclusions}
The substitution of nonmagnetic impurities into the layered
cuprates provides useful information about the many-body physics
of these materials. In particular, the nonmagnetic impurities
constitute a perturbation in the density channel, and the Knight
shift measurements determine the real-space resolved response of
the system to this perturbation. In the NMR experiments on $^7$Li,
$^{89}$Y and $^{63}$Cu nuclei in YBa$_2$Cu$_3$O$_{6+x}$ with
nonmagnetic impurities, it has been found that, when a uniform
magnetic field is applied, the electronic spins in CuO$_2$ planes
polarize to form a magnetization with an oscillatory pattern
\cite{Mahajan94,Bobroff,Mendels,Mahajan2000,MacFarlane,Julien}. In
order to understand the origin of this behavior, here QMC data
have been presented for one nonmagnetic impurity in the 2D Hubbard
model. These QMC data show that, as the temperature is lowered and
the AF fluctuations grow, giant oscillations develop in the Knight
shift response around the impurity. The pattern of these
oscillations and of the electron density were discussed here. In
addition, the effective impurity potential extracted from the QMC
data was shown. Since these results are restricted to high
temperatures, it is not possible to compare directly with the NMR
experiments. For this reason, the QMC data were compared with the
results of the previous RPA analysis of the NMR data. It was found
that, as $T$ decreases, the patterns which develop in the Knight
shift response and the local electron density are in agreement
with those found in the RPA analysis by fitting the experimental
data.

In spite of these, a number of open questions remain. For
instance, it is not known whether, within the 2D Hubbard model
with only nearest-neighbor hopping, a fit to the perfect
Curie-Weiss $T$ dependence of the $^7$Li Knight shift will be
obtained at low $T$ between 100K and 400K. Perhaps, a
second-near-neighbor hopping or other terms are required for
describing the pure material. It is also possible that the bare
impurity potential is more involved than the simple form of
Eq.~(\ref{Vimp}) used here. Nevertheless, the QMC data presented
here clearly show the role of the AF fluctuations in producing the
giant oscillations in the Knight shift response.

\acknowledgments

The author gratefully acknowledges helpful discussions with H.
Alloul, J. Bobroff, W. Hanke and M. Imada. The author also thanks
J. Bobroff for helpful comments about the manuscript. This work
was supported by the Turkish Academy of Sciences through the GEBIP
program (EA-T\"{U}BA-GEBIP/2001-1-1). The author thanks the
Institute for Solid State Physics at the University of Tokyo and
the Institute for Theoretical Physics at the University of
W\"{u}rzburg for their hospitality. The numerical computations
reported in this paper were performed at the Center for
Information Technology at Ko\c{c} University.

%\newpage

\newpage

\begin{figure}
\begin{center}
\leavevmode \epsfxsize=7.5cm \epsfysize=6.59cm \epsffile[100 170
550 580]{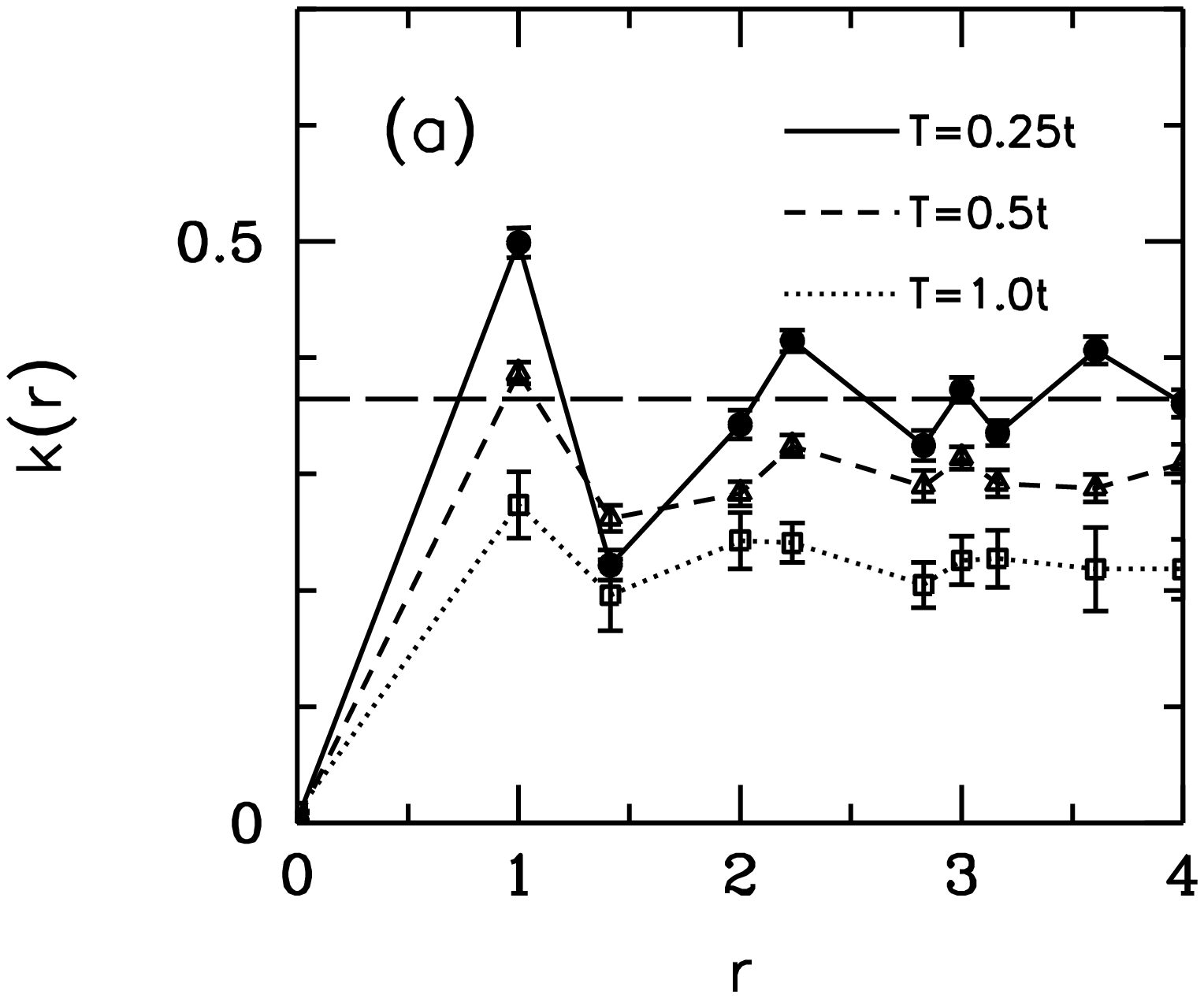}
\leavevmode \epsfxsize=7.5cm \epsfysize=6.59cm
\epsffile[100 170 550 580]{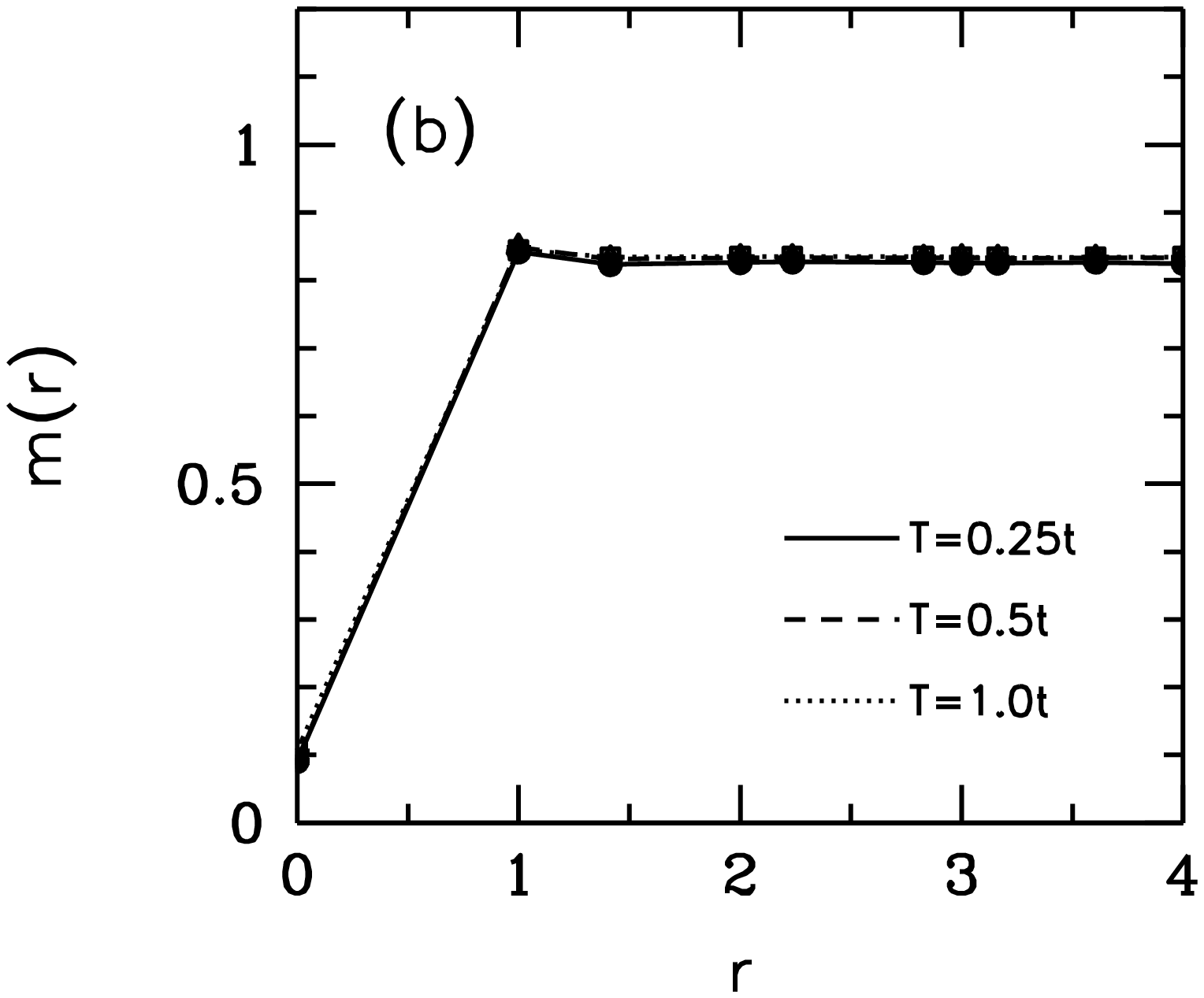}
\end{center}
\caption{ (a) Knight shift response function $k({\bf r})$ versus
$r=|{\bf r}|$ away from the impurity at various temperatures.
Here, $r$ is plotted in units of the lattice spacing. These
results are for $U=4t$ and $\langle n\rangle =0.875$ on an
$8\times 8$ lattice. The long-dashed horizontal line denotes the
value of $k({\bf r})$ for the pure system at $T=0.25t$. (b) Local
magnetic moment $m({\bf r})=\sqrt{\langle (m^z({\bf
r}))^2\rangle}$ versus $r$ for the same parameters as in (a). }
\label{fig1}
\end{figure}

\newpage

\begin{figure}
\begin{center}
\leavevmode \epsfxsize=7.5cm \epsfysize=6.59cm \epsffile[100 170
550 580]{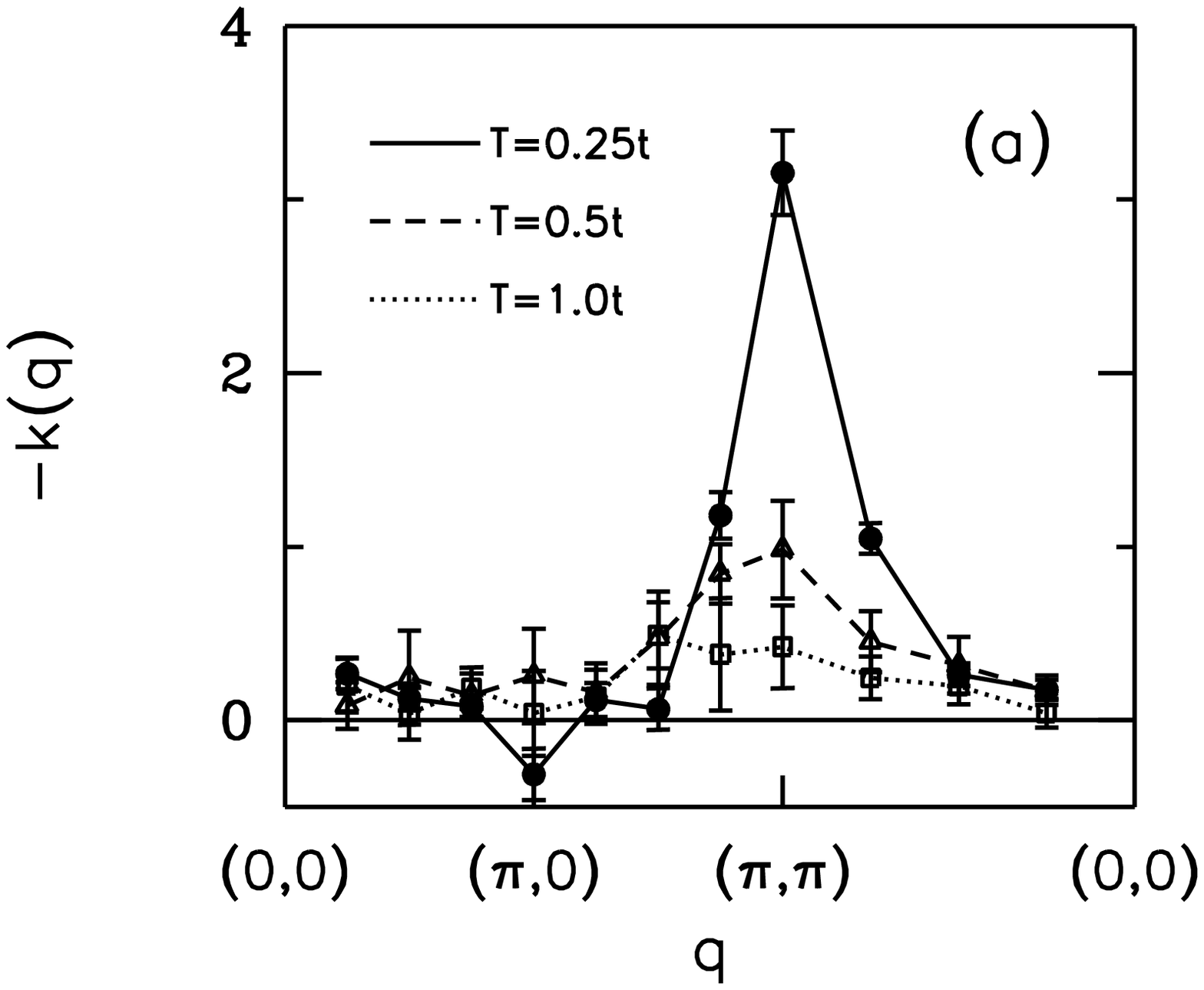}
\leavevmode \epsfxsize=7.5cm \epsfysize=6.59cm
\epsffile[100 170 550 580]{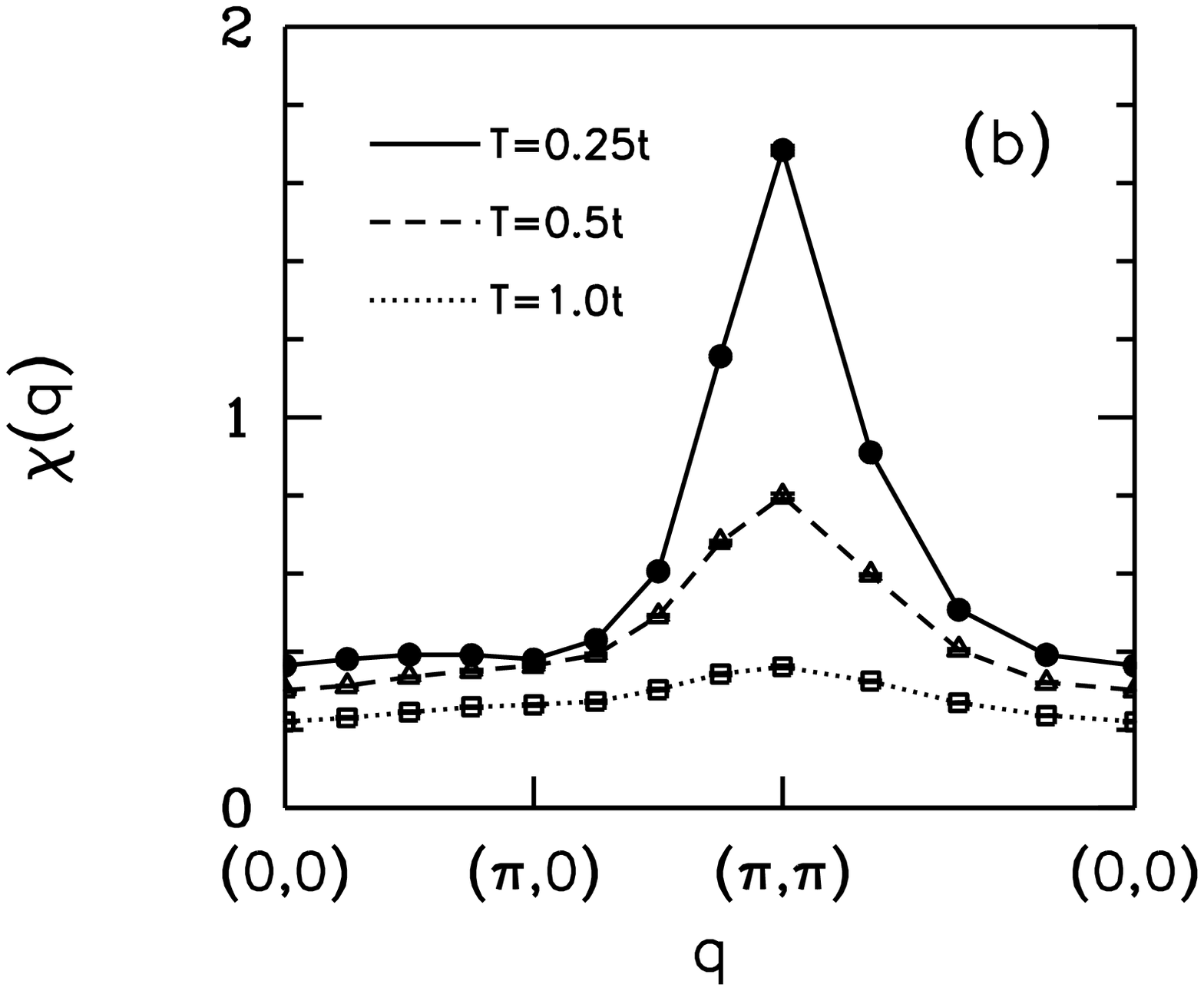}
\end{center}
\caption{ (a) Wavevector dependence of the Knight shift response
function $k({\bf q})$ along various cuts in the Brillouin zone.
(b) Diagonal magnetic susceptibility $\chi({\bf q})=\chi({\bf
q},{\bf q},i\omega_m=0)$ versus ${\bf q}$. These results are for
$U=4t$, $\langle n\rangle=0.875$ and an $8\times 8$ lattice at
various temperatures. } \label{fig2}
\end{figure}

\newpage

\begin{figure}
\begin{center}
\leavevmode \epsfxsize=7.5cm \epsfysize=6.59cm \epsffile[100 180
550 630]{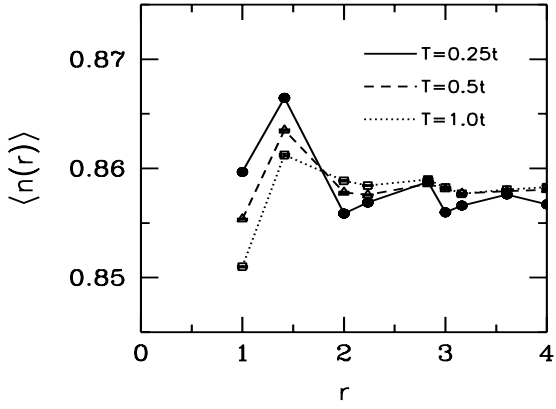}
\end{center}
\caption{ Site-dependent electron density $\langle n({\bf
r})\rangle$ versus $r$ plotted in the same way as in Fig.~1. These
data are for $U=4t$, $\langle n\rangle=0.875$ and an $8\times 8$
lattice at various temperatures. } \label{fig3}
\end{figure}

\newpage

\begin{figure}
\begin{center}
\leavevmode \epsfxsize=7.5cm \epsfysize=6.59cm \epsffile[100 170
550 580]{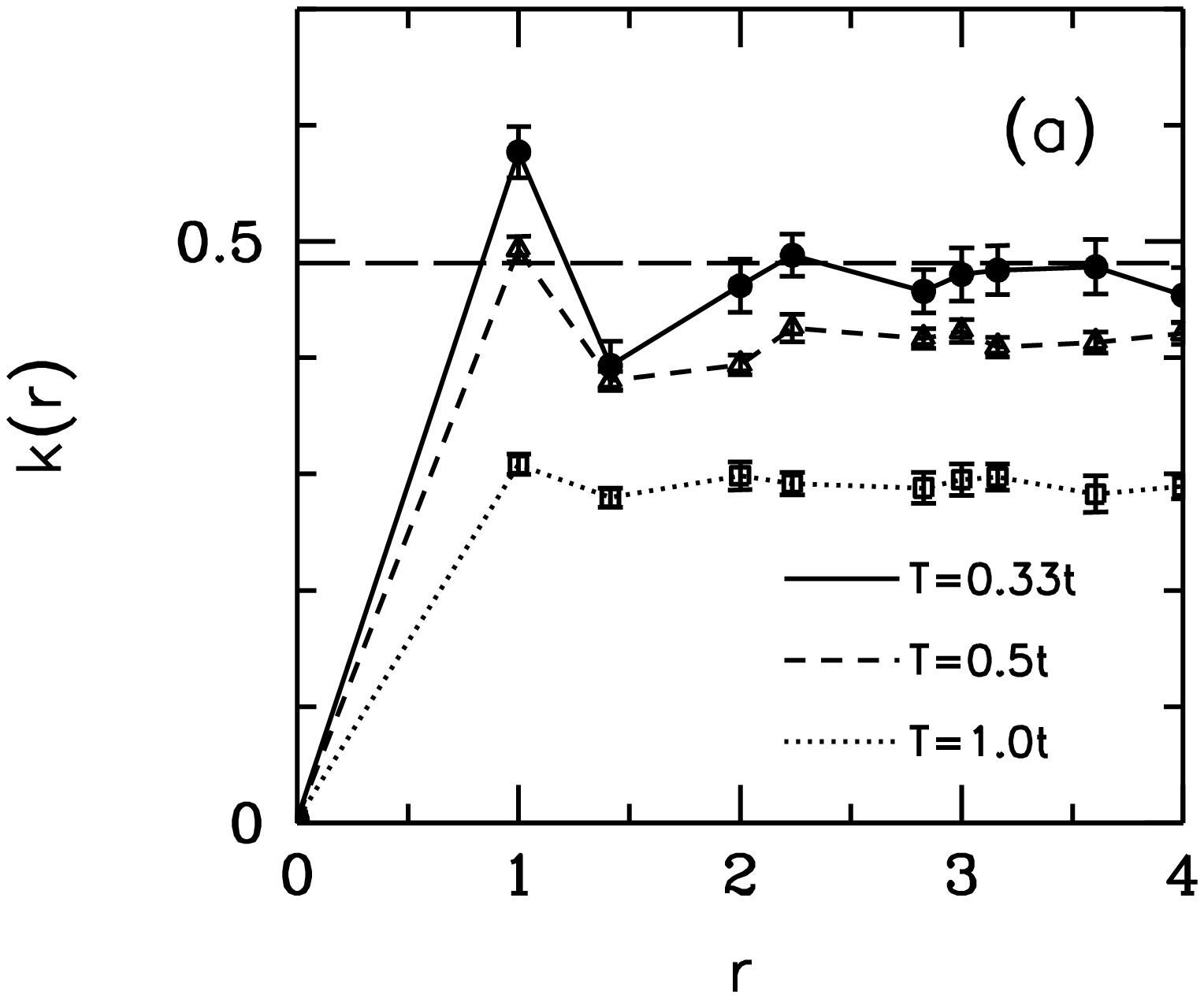}
\leavevmode\epsfxsize=7.5cm\epsfysize=6.59cm\epsffile[100 170 550
580]{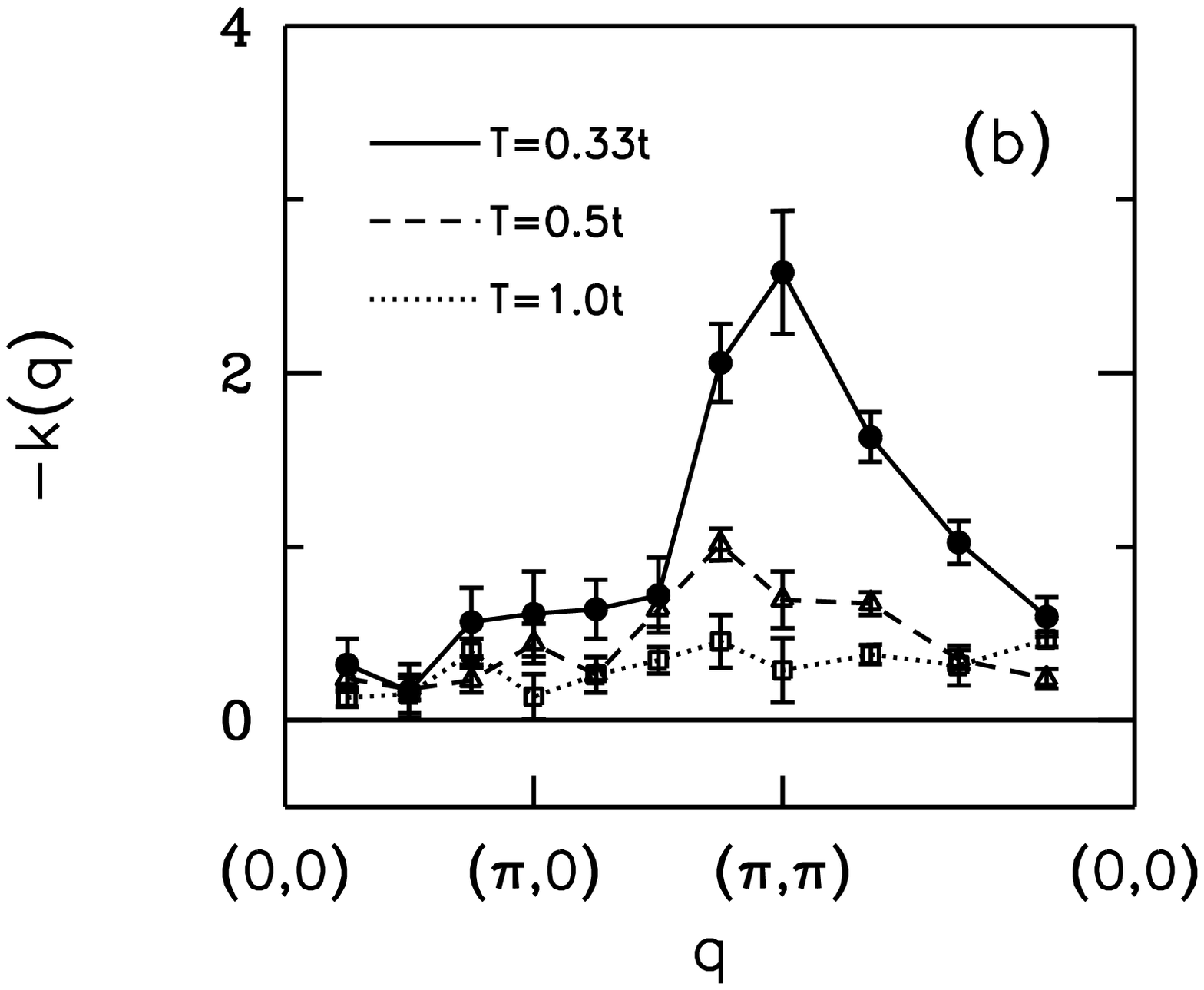}
\end{center}
\begin{center}
\leavevmode \epsfxsize=7.5cm \epsfysize=6.59cm
\epsffile[100 170 550 580]{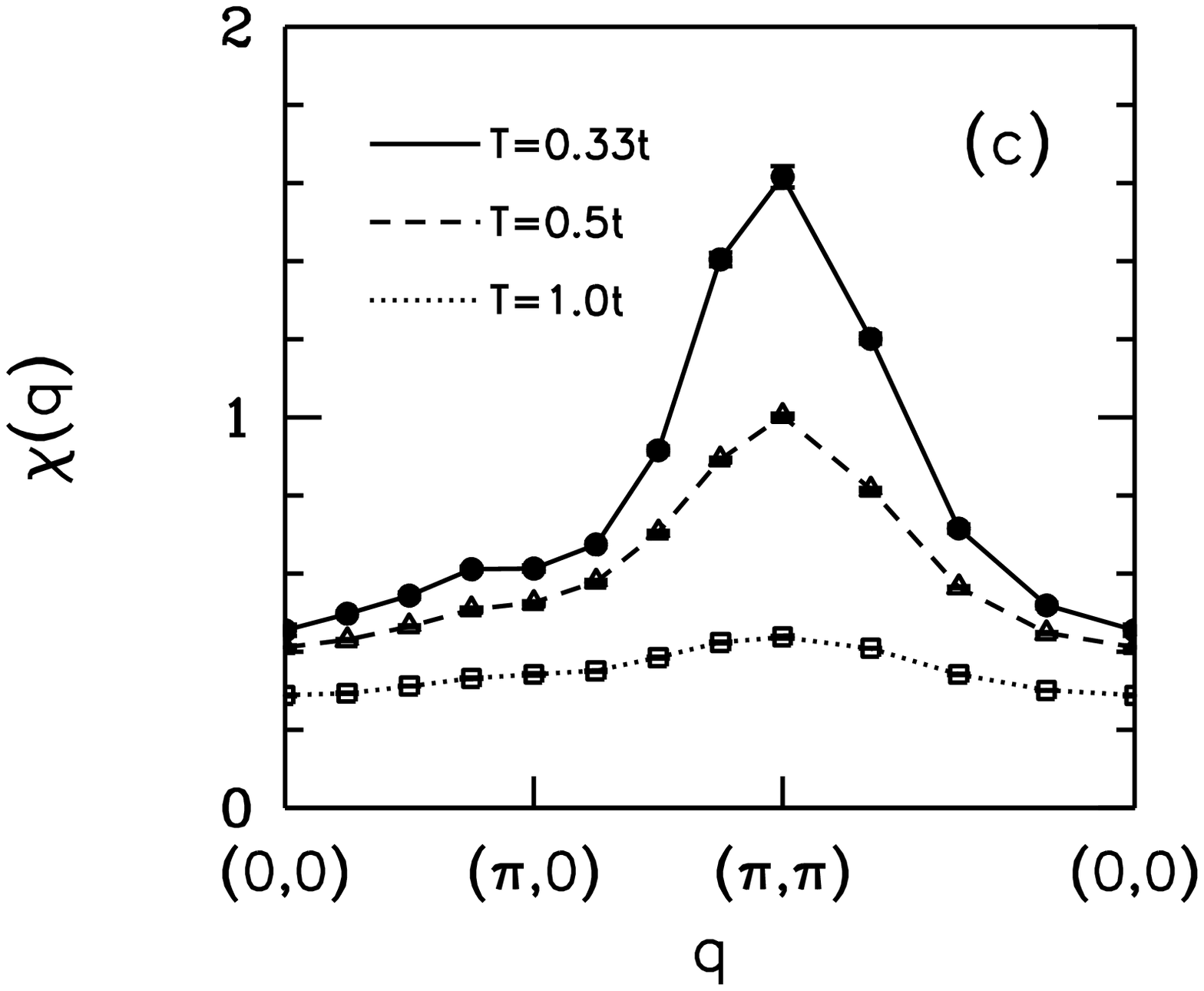} \leavevmode \epsfxsize=7.5cm
\epsfysize=6.59cm \epsffile[100 170 550 580]{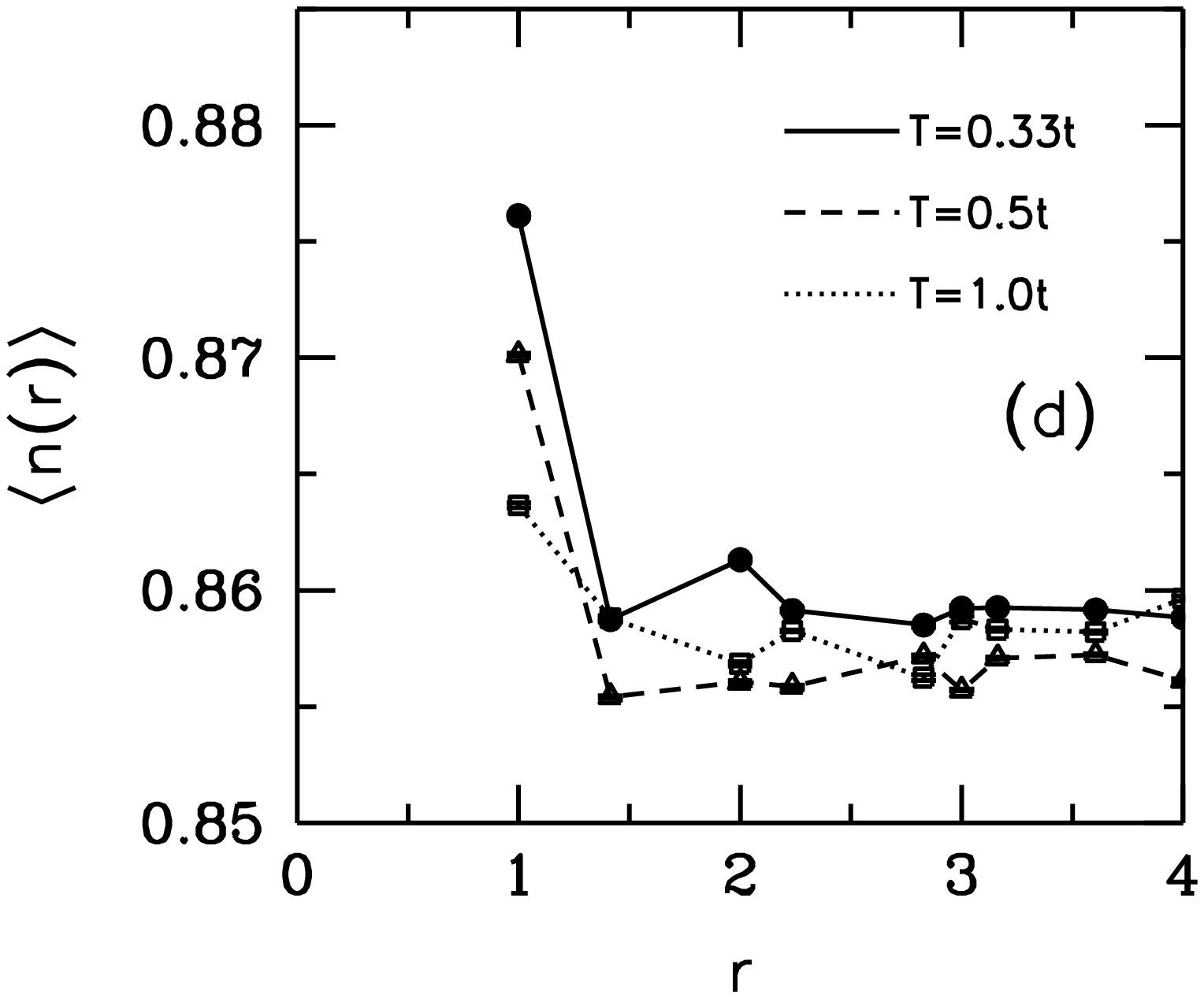}
\end{center}
\caption{ (a) $k({\bf r})$ versus $r$, (b) $-k({\bf q})$ versus
${\bf q}$, (c) $\chi({\bf q})$ versus ${\bf q}$ and (d) $\langle
n({\bf r})\rangle$ versus $r$ for $U=8t$, $\langle n({\bf
r})\rangle=0.875$ and an $8\times 8$ lattice at various
temperatures. In (a), the long-dashed horizontal line denotes the
value of $k({\bf r})$ for the pure system at $T=0.33t$. }
\label{fig4}
\end{figure}

\newpage

\begin{figure}
\begin{center}
\leavevmode \epsfxsize=7.5cm \epsfysize=6.59cm \epsffile[100 170
550 580]{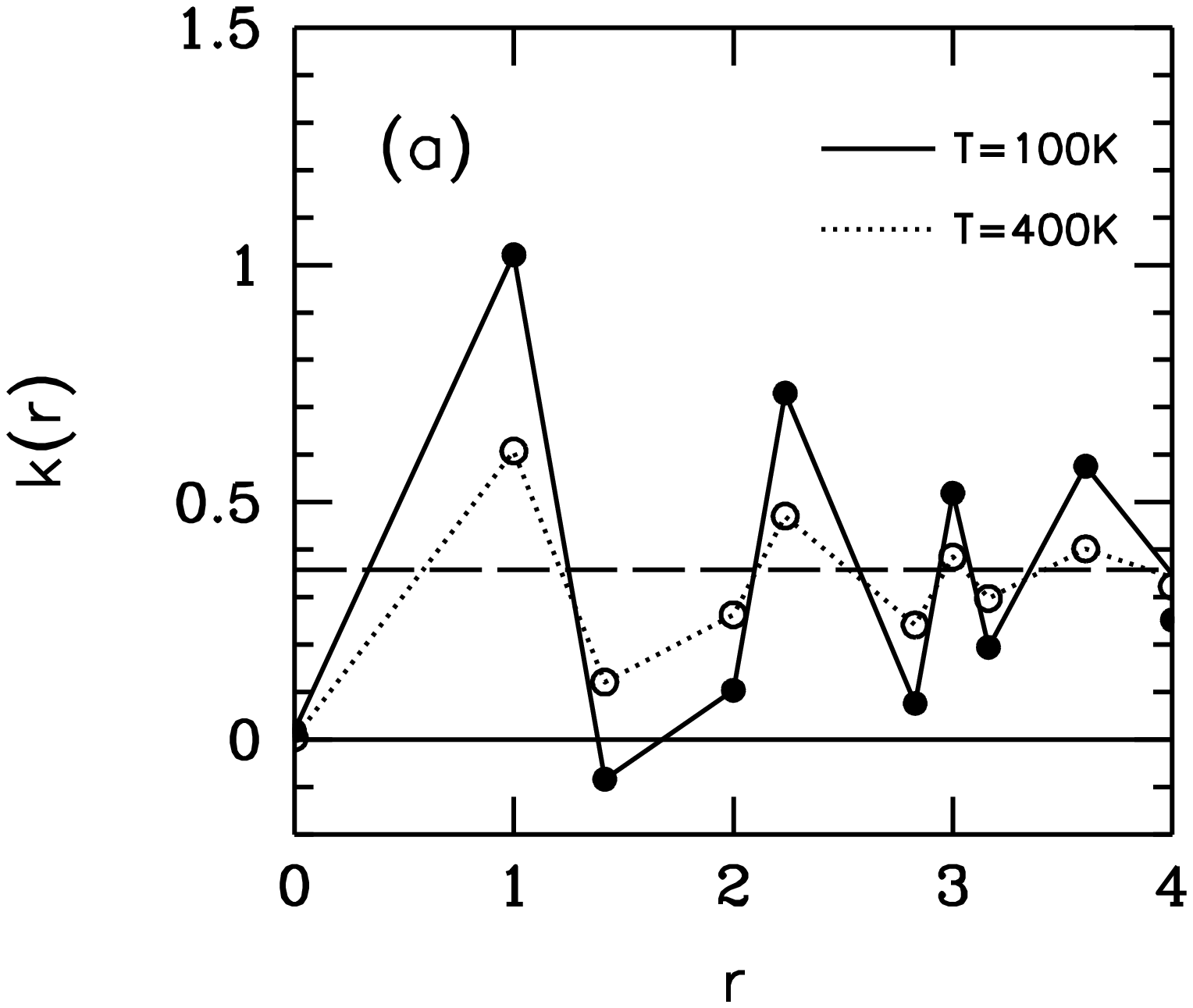} \leavevmode \epsfxsize=7.5cm \epsfysize=6.59cm
\epsffile[100 170 550 580]{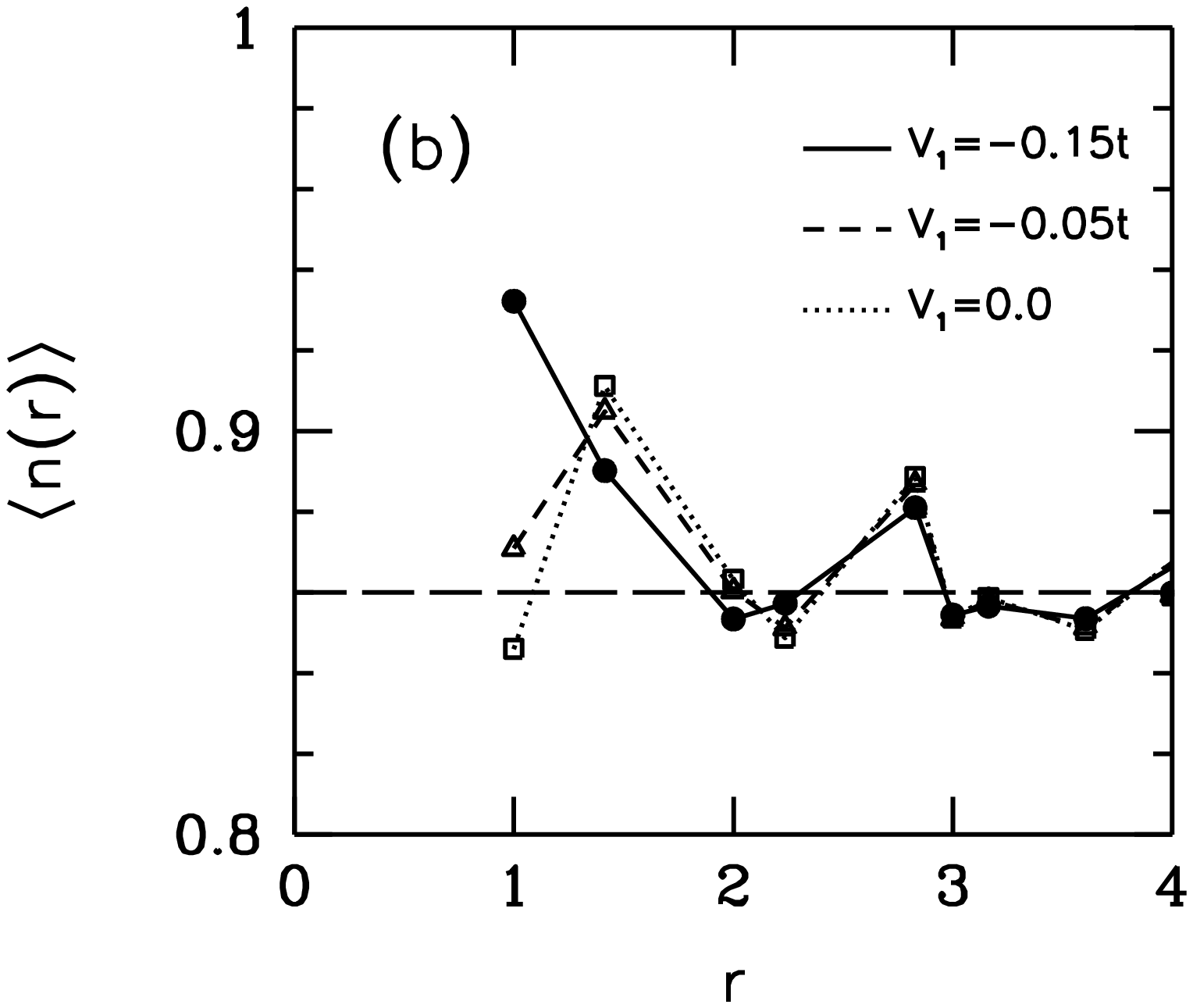}
\end{center}
\caption{ (a) Knight shift response function $k({\bf r})$ versus
$r$ at 100K and 400K obtained for the optimally doped YBCO with Li
impurities. This result was obtained within an RPA analysis of the
experimental data by using $V_1=-0.15t$. Here, $V_1$ is the value
of the effective impurity potential at the sites neighboring the
impurity. (b) $\langle n({\bf r})\rangle$ versus $r$ obtained from
the same RPA calculations at 100K for different values of $V_1$.
In these figures, the long-dashed horizontal lines represent the
values of $k({\bf r})$ and $\langle n({\bf r})\rangle$ for the
pure system at $T=100$K. } \label{fig5}
\end{figure}

\newpage

\begin{figure}
\begin{center}
\leavevmode \epsfxsize=7.5cm \epsfysize=6.59cm \epsffile[100 170
550 580]{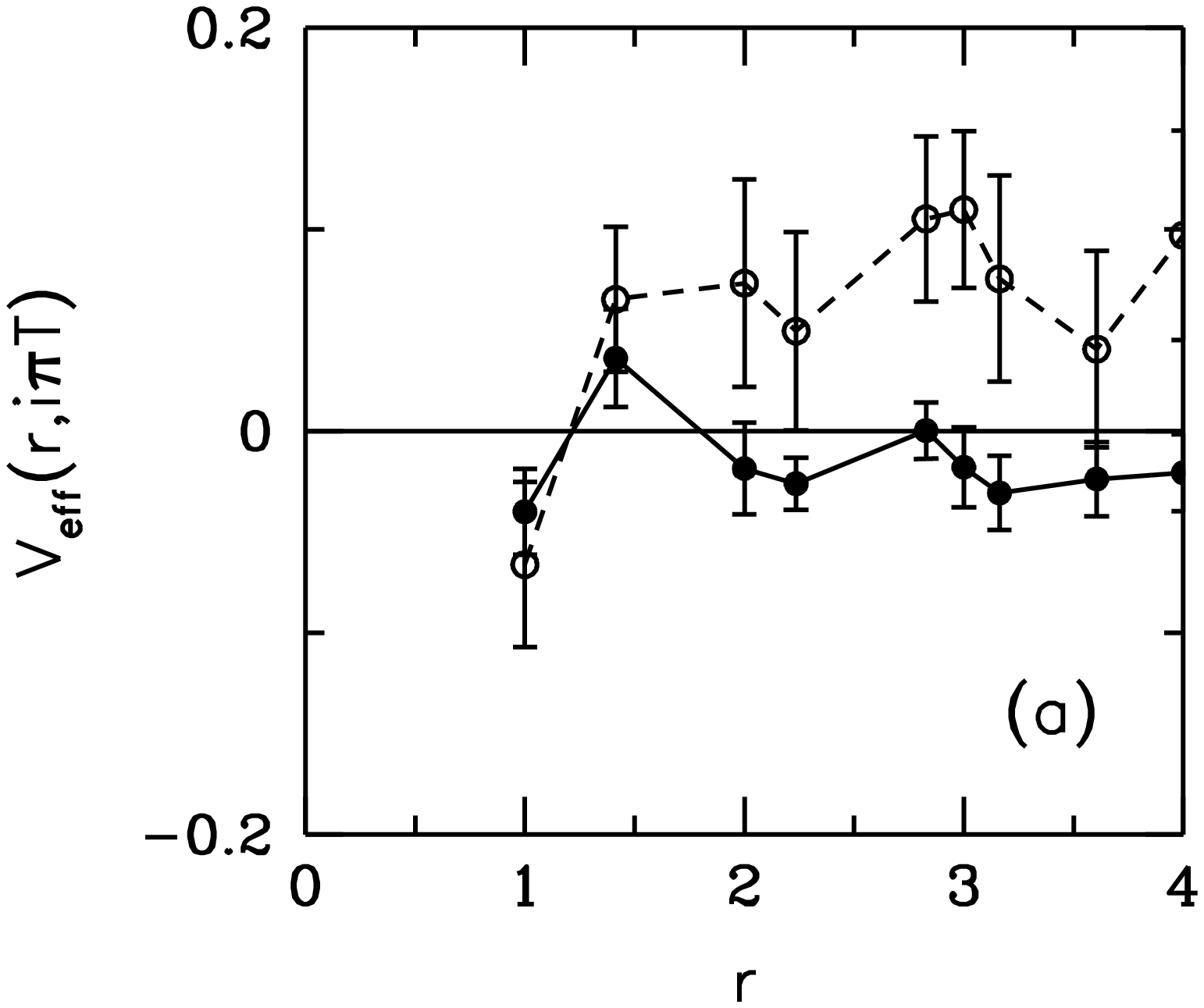} \leavevmode \epsfxsize=7.5cm \epsfysize=6.59cm
\epsffile[100 170 550 580]{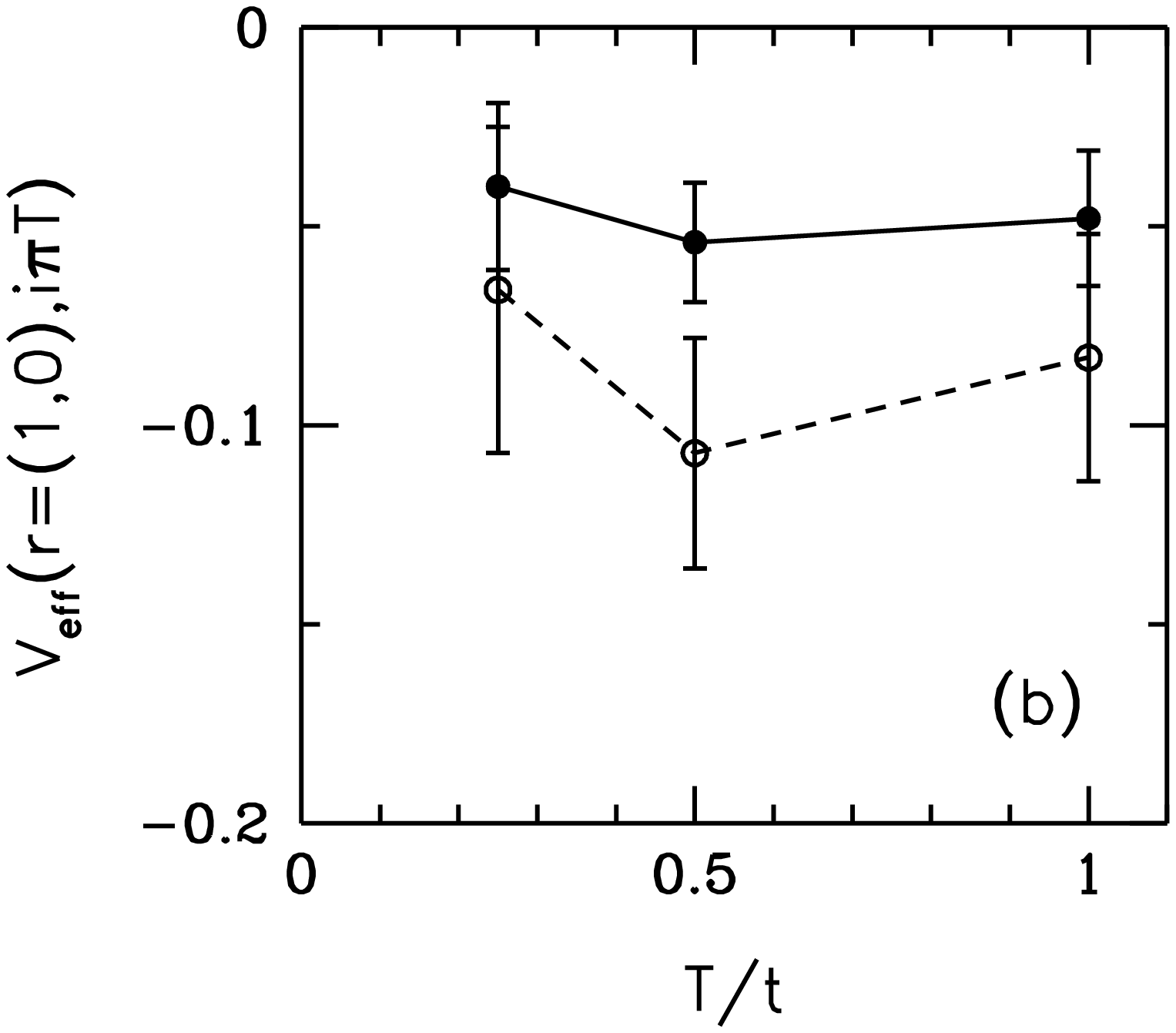}
\end{center}
\caption{ (a) Effective impurity potential $V_{eff}({\bf r},i\pi
T)$ versus $r$ at $T=0.25t$. (b) $V_{eff}({\bf r},i\pi  T)$ at
${\bf r}=(1,0)$ versus $T$. Here, the filled circles denote the
real part of $V_{eff}$, and the open circles denote the imaginary
part. These results are for $U=4t$ and $\langle n\rangle =0.875$.}
\label{fig6}
\end{figure}

\newpage

\begin{figure}
\begin{center}
\leavevmode \epsfxsize=7.5cm \epsfysize=6.59cm \epsffile[100 170
550 580]{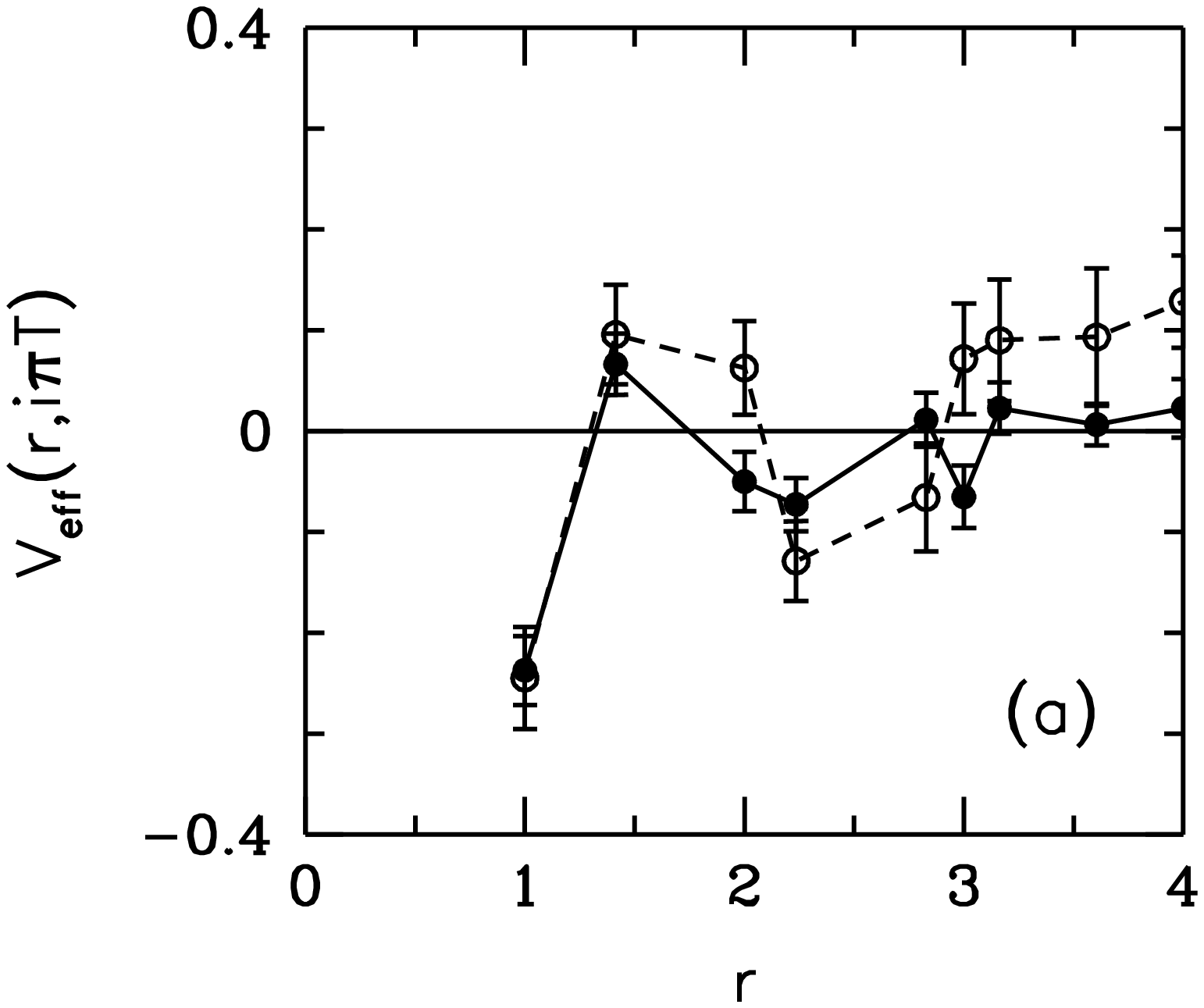} \leavevmode \epsfxsize=7.5cm \epsfysize=6.59cm
\epsffile[100 170 550 580]{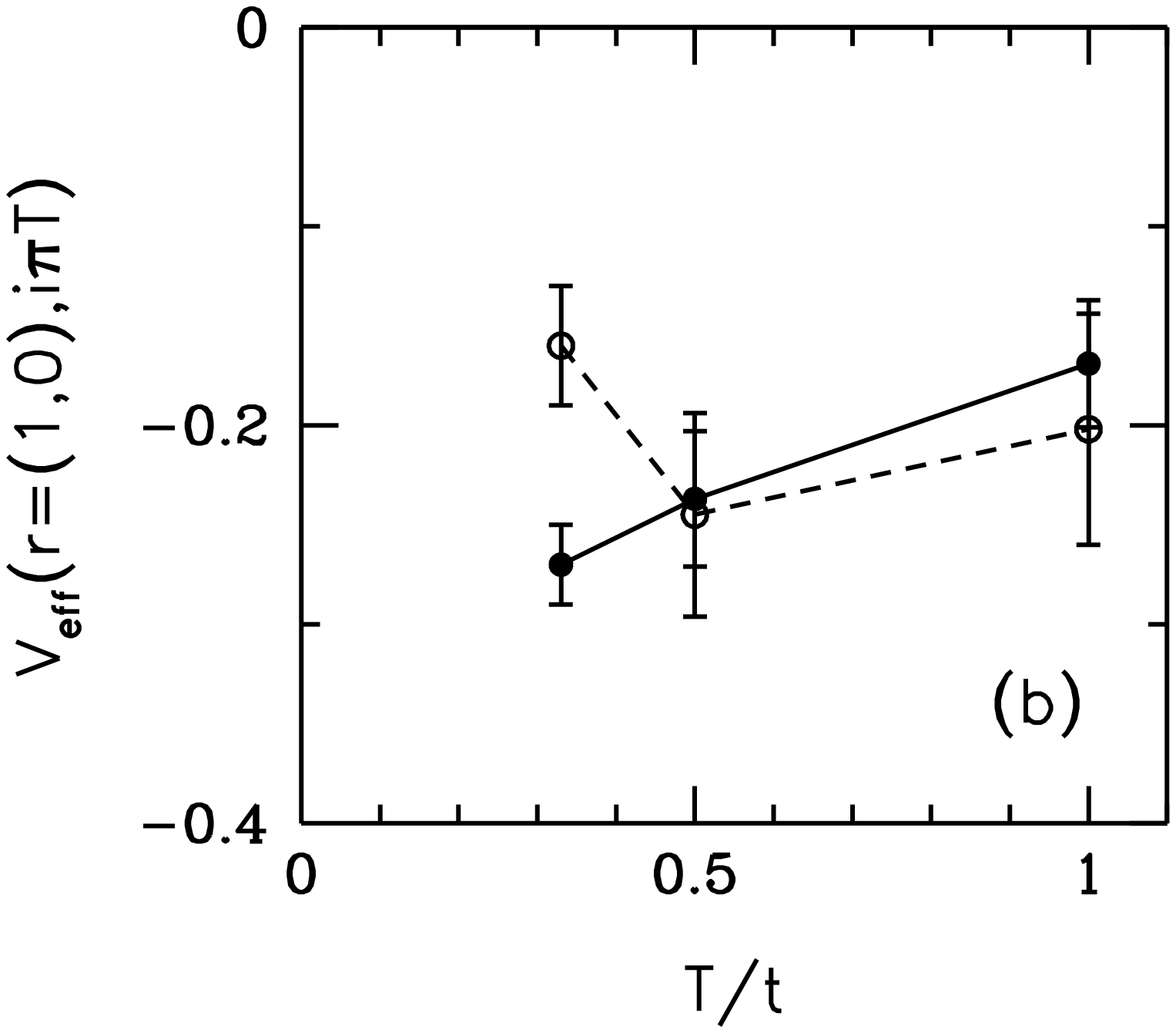}
\end{center}
\caption{ (a) Effective impurity potential $V_{eff}({\bf r},i\pi
T)$ versus $r$ at $T=0.5t$. (b) $V_{eff}({\bf r},i\pi  T)$ at
${\bf r}=(1,0)$ versus $T$. Here, the filled circles denote the
real part of $V_{eff}$, and the open circles denote the imaginary
part. These results are for $U=8t$ and $\langle n\rangle =0.875$.
} \label{fig7}
\end{figure}

\end{document}